\documentstyle[aps,prd,epsf,floats,amsmath]{revtex}

\newcommand{\bk}{{\bf k}}
\newcommand{\bq}{{\bf q}}
\newcommand{\bzero}{{\bf 0}}

\newcommand{\STAR}{{
\begin{picture}(8,8)
 \put(0,0){+}
 \put(0,0){$\times$}
\end{picture}}}

\newcommand{\Z}{{\sf
\begin{picture}(10,10)
 \put(0,0){Z}
 \put(1.5,0){Z}
\end{picture}}}

\title{Scalar density fluctuation at critical end point in NJL model}
\author{H. Fujii\footnote{Email: hfujii@phys.c.u-tokyo.ac.jp} }
\address{%
Institute of Physics, University of Tokyo\\
3-8-1 Komaba, Meguro, Tokyo 153-8902, Japan}
\date{\today}

\begin{document}
\draft
\maketitle

\begin{abstract}
Soft mode near the critical end point in the phase diagram of
two-flavor Nambu--Jona-Lasinio (NJL) model is investigated
within the leading $1/N_c$ approximation with $N_c$ being the
number of the colors.
 It is  explicitly shown by studying the
spectral function of the scalar channel that the relevant soft mode is
the scalar density fluctuation, which is coupled with the quark number
density, while the sigma meson mode stays massive.  
\end{abstract}


\section{Introduction}

Existence of critical end point (CEP)
in the QCD phase diagram is recently suggested in several works
\cite{AY89,GGP94,BR99,HJSSV98,FK02},
and its implications to
the phenomena in the heavy ion physics are intensively discussed
in order to identify its location  experimentally
\cite{RS98,RS99,SG99,BR00,SMMR01,BPSS01,HI03,KF03,HS03}.

The critical end point is the end point of the 
phase boundary determined by the first order phase transition.
The nature of this point has been discussed so far
mainly from the viewpoint of the chiral  symmetry.
The order parameter is the 
scalar quark condensate, $\langle \bar q q \rangle \sim \sigma$,
which strictly vanishes in the Wigner phase in the chiral limit.
At the CEP in question
 the phase transition necessarily becomes of second order
even with the non-zero current quark mass $m \ne 0$,
and the curvature of the
free energy w.r.t.\ the
scalar field should vanish. This fact immediately implies
the emergence of a gapless excitation in the scalar channel.
The existence of this second-order transition, however,
is accidental in the sense that no symmetry
change of the ground state is accompanied with this transition.
In fact the finite current quark mass,
which breaks the chiral symmetry explicitly, results in 
the non-vanishing scalar condensate at any point 
in the plane of the temperature and 
the quark chemical potential ($T-\mu$),
in contrast to the case with $m=0$. 
Only its fluctuation 
around the equilibrium value becomes unstable at the CEP. 
This ``accidental'' instability is not
related to any symmetry of the system. 

Even in the chiral symmetric
world with $m=0$, presence of the tricritical point (TCP),
which is the counterpart of the CEP, is not necessarily required
by the spontaneous breaking of the chiral symmetry.
The change of the symmetry realization of the system
constitutes a boundary line in the phase diagram, on which
the first-order or the second-order transition occurs.
The soft mode for the chiral breaking is known as sigma meson,
and it becomes gapless on the second-order line so as to recover
the chiral symmetry together with the Goldstone pion. 
The TCP exists as a point separating the boundary line
into the segments of the first-order and the second-order
transitions, depending on the parameters $T$ and $\mu$.
What kind of singular behavior will appear additionally
at this TCP, where the sigma mode is already gapless?

The most natural expectation may be that the density fluctuation 
becomes soft near the CEP because 
it is the end point of the
first order transition which accompanies the density gap
caused by the strong attraction between
the quarks. 
The aim of this paper is to prove this scenario
explicitly within the NJL model\cite{HK94,SPK92} in the leading $1/N_c$
approximation, and to show that the nature
of the CEP is rather similar to the liquid--gas
critical point than the chiral transition.
The density fluctuation discussed here
will be the collective motion of the particle-hole-like excitations.
One should note that this density fluctuation has a space--like dispersion
relation from the kinematical reason. Thus,
we will be forced to investigate the mode spectrum
with this kinematics.
If this is the case, some of the implications discussed before
in the literature\cite{RS98,SMMR01,KF03}
must be reconsidered in relation to the heavy ion experiments;
at the CEP the sigma meson, which is the chiral partner of the pion, 
can remain massive. One should study the observables
distinctive of the space-like density fluctuations,
rather than the particle production processes with the time--like kinematics.

The nature of the phase transition is insensitive to the microscopic 
structure of the system. The most important ingredients which
determine the nature of the transition are the dimension of the
system, the symmetry of the interactions or
 the dimension of the order parameter(s), and the range
of the interactions. If the study 
is extended to include
the dynamical nature,
these universarity classes for the static properties
split into subclasses depending on the existence of
the conservation law(s) and mode-mode coupling(s), and so on
\cite{HH77}.
It may be, therefore, perilous to utilize (e.g.) the Ising model results
in order to speculate the dynamical 
nature of the QCD phase transition. With this caution
in mind we employ
a concrete model of the quarks in this paper to study the
dynamical properties of the phase transition at the CEP.

We use the NJL model here as a simple model which  can
describe the CEP as well as
the chiral phase transition of the quark system.
One of the advantages of the use of the NJL model is
that 
the bosonic collective modes of the system are easily and explicitly
calculated.
Our main conclusions are expected to hold irrespective of
the details of the model employed here. 
We should still keep cautious, however, because
this model lacks
the long-range nature of the color interactions, and the gluonic 
degrees of freedom at all.
In this paper we restrict our discussion only for the second order 
phase transition at the CEP ($m\ne0$). More complete analysis 
treating the TCP, and the vector and entropy density fluctuations,
will be reported in the forthcoming paper\cite{FOO03}
in the NJL model 
as well as in the linear sigma model.

This paper is organized as follows. In \S\ref{sec:2}
the behavior of the susceptibilities in the NJL phase transition 
is discussed. In \S\ref{sec:3} the spectral function of the scalar mode
is investigated in some details and the relevant slow mode is explicitly 
identified as the scalar density fluctuation.
\S\ref{sec:4} is devoted to
summary.

\section{Susceptibilities in the NJL phase transition}
\label{sec:2}

We use the simplest version of the NJL model 
with two flavors ($N_f=2$)\cite{NJL}
whose laglangian is, with obvious notations,
\begin{equation}
{\cal L} = \bar q  (i \!\!\! \not\! \partial -m) q +g [
(\bar q q)^2 + ( \bar q i \gamma_5 \tau^a q)^2].
\end{equation}
After introducing the auxiliary fields $\sigma$ and $\pi$
for $\bar q q$ and $\bar q i \gamma_5 \tau^a q$ in a standard way,
the system pressure at finite temperature $T$ and finite quark 
chemical potential $\mu$ is written,
within the mean--field approximation
with $\sigma$=const and $\pi$=0, as
\begin{eqnarray}
{1 \over \nu} P(T,\mu,m;\sigma)&=&
 \int {d^3 k\over (2\pi)^3}
[E - T \ln (1-n_-) -T \ln (1-n_+)]
-{1\over 4}{1\over g\nu}\sigma^2,
\end{eqnarray}
where $n_{\pm}=(e^{\beta(E \mp \mu)}+1)^{-1}$, $E=\sqrt{M^2+\bk^2}$,
$M=m-\sigma$, and $\nu=2 N_f N_c$ is the number of
the quark species. 
This expression corresponds to the leading $1/N_c$ 
approximation with $g N_c$ being $O(1)$,
though we set $N_c=3$ in the following.
From the pressure $P(T,\mu,m;\sigma)$ we can calculate all the
thermodynamical quantities.  

The ground state at ($T,\mu$) satisfies the stationary condition
for $\sigma$, 
which yields the gap (saddle point) equation
\begin{equation}
\int {d^3 k\over (2\pi)^3}{M\over E}(1-n_- -n_+)
+{1 \over 2}{1 \over g \nu}\sigma=0.
\end{equation}
We define the curvature of the potential w.r.t.\ the scalar field
at the saddle point
\begin{eqnarray}
-{\partial^2 P \over \partial \sigma^2} &\equiv& K
\nonumber  \\
&=& {1 \over 2 g}-
\nu \int {d^3 k\over (2\pi)^3}{1\over E}(1-n_- -n_+)
+
 \nu \int {d^3 k\over (2\pi)^3}
{M^2 \over E^2}
\left ({1-n_- - n_+ \over E}+
n'_- + n'_+ \right )
\nonumber  \\
&=& {1 \over 2 g}-\chi_{mm}^{(0)}
 \ge 0
\label{eq:curvature}
\end{eqnarray}
with $n'_\pm = \partial n_\pm / \partial E$. 
The last inequality must be satisfied according to thermodynamical stability.
The occurrence of the second order phase transition is signaled by the
vanishing curvature, $K=0$.
Furthermore, we notice from Eq.~(\ref{eq:curvature})
together with the gap equation, 
that $K \sim M^2 \to 0$ 
at the second order chiral transition approached 
from the lower temperature or
density. 
As we shall see later, $K$ is proportional to the inverse propagator
of the static scalar mode in the long wave length limit.

\begin{figure}[tb]
\centerline{\epsfxsize=0.5\textwidth\epsffile{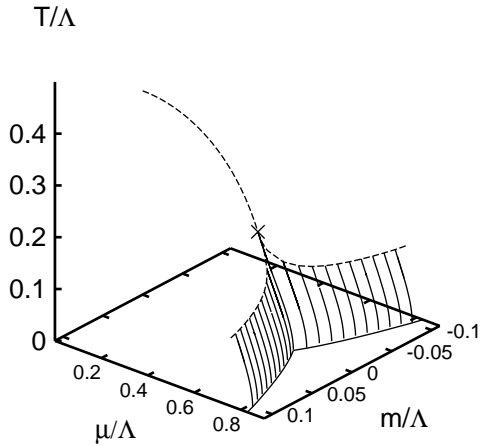}}
\caption{Calculated phase diagram of the NJL model. The transition line of the
second order is drawn in the dashed line. The phase boundary of 
the first order transition constitutes a surface shown by hatch.
The cross indicates the tricritical point, where three lines of second
order transition meet.} 
\label{fig:phase}
\end{figure}

The susceptibilities $\chi_{ab}=\partial ^2 P/\partial a \partial b$
($a,b=T,\mu, m$) are very important to characterize the nature of
the phase transition
\cite{TK91,AG91}.
In our approximation these susceptibilities have a simple common
structure as
\begin{equation}
\chi_{ab} = \chi^{(0)}_{a m}{1 \over K}\chi^{(0)}_{m b }+\chi^{(0)}_{ab}, 
\label{eq:susceptibility}
\end{equation}
where $\chi_{ab}^{(0)}$ are the non-singular parts of the
susceptibilities and equal to those of the free quark gas
of mass $M$. 
From this expression we find that all divergences at the second order
phase  transition originate from the flatness of the potential
curvature  $K\to 0$.
One should note here that
the mixing of the susceptibilities with the scalar mode
$\chi_{am}^{(0)}$ ($a=T,\mu$) is proportional to $M$.
Thus, at the chiral phase transition where $m=0$,
the would-be singular parts of $\chi_{ab}$ ($a,b=T,\mu$) involving
the factor $1/K$, are less singular than $\chi_{mm}$.
Actually they remain finite (vanish)
if the transition point ($T_c, \mu_c$)
is approached from the broken (symmetric) phase
because the constituent quark mass $M$ vanishes on the boundary
and in the symmetric phase. 
On the other hand, at the CEP ($m\ne 0$), $M$ stays finite and
the critical behavior of these susceptibilities is essentially 
the same as $\chi_{mm}$. This argument is consistent with 
the analysis based on the general properties of 
the Landau free energy with the tricritical point
\cite{HI03}, as it should be.

The discussion can be extended a little bit further by changing the
thermodynamical variables. Let us choose the quark number density
$\rho_q$ as an independent variable instead of $\mu$. 
Then the scalar susceptibility with fixed quark number density is
related to the susceptibilities given in Eq.~(\ref{eq:susceptibility})
via
\begin{equation}
\chi_{mm}(T,\rho_q) =
{1 \over \chi_{\mu\mu}}(\chi_{mm}\chi_{\mu\mu}-\chi_{m\mu}^2)
=\frac{
(\chi^{(0)}_{mm}\chi^{(0)}_{\mu\mu}-{\chi^{(0)}_{m\mu}}^2)
(1+{1 \over K}\chi^{(0)}_{mm})}
{\chi^{(0)}_{\mu\mu}+\chi^{(0)}_{\mu m}\frac{1}{K}\chi^{(0)}_{m\mu}}.
\label{eq:fixedrho}
\end{equation}
We find that the $K^{-2}$ term is canceled out in the numerator.
At the chiral phase transition we already pointed out 
that $\chi_{\mu\mu} \to $ constant, and then
the scalar susceptibility with the fixed quark number density
$\chi_{mm}(T,\rho_q)$ diverges in the same way as
$\chi_{mm}$. There
the vector susceptibility is decoupled
from the scalar one as $\chi^{(0)}_{m\mu}\sim M \to 0$.
In case of the CEP which we are concerned about, however, 
$\chi^{(0)}_{m\mu}$ remains finite, and the denominator 
and the numerator in Eq.~(\ref{eq:fixedrho})
diverges in the same way as $1/K$, which results in the
finite scalar susceptibility. This demonstrates that 
the coupling with the
number density fluctuation is essential 
for the divergence accompanied by the CEP. 
Similarly the thermal susceptibility with fixed quark number density,
$\chi_{TT}(T,\rho_q)
=(\chi_{TT}\chi_{\mu\mu}-\chi_{\mu T}^2)/\chi_{\mu   \mu}$,
does not diverge at the CEP, but $\chi_{TT}$ in
Eq.~(\ref{eq:susceptibility}) does.

This situation is completely analogous to the thermal susceptibility
or the specific heat in the liquid--gas phase transition, where
the density gap between two phases
is identified as the order parameter.
The specific heat at constant pressure and the isothermal compressibility
diverge like $|T-T_c|^{-\gamma}$ at the critical point. 
Once we keep the number (entropy) density fixed,
however, the specific heat at constant volume (the isentropic
compressibility) remains finite or diverges only weakly like
 $|T-T_c|^{-\alpha}$)\cite{HES77}.

The NJL model has two fundamental parameters: 
the coupling constant $g$ and the 
cutoff $\Lambda$. The ``physical values'' of these parameters are
usually fixed
so as to reproduce the 
chiral quark condensate $\langle \bar qq \rangle$ and the pion decay
constant $f_\pi$. 
Taking the three--momentum cutoff scheme, 
the critical value $g_c \Lambda^2$ for the spontaneous symmetry breaking in the
vacuum is readily found as $g_c \Lambda^2={2 \pi^2 \over \nu}=1.64\cdots$
in the chiral limit ($m=0$).
There is also a critical value for the appearance of the
TCP, which value is obtained as
 $g_{TCP} \Lambda^2={2 \pi^2 \over \nu}{1 \over 1-e^{-2}}=1.90\cdots$
from the condition that the $\sigma^2$-- and 
$\sigma^4$--terms simultaneously vanish in the pressure
$P(T,\mu,0;\sigma)$.
It is interesting to note that the TCP must appear 
with the finite chemical potential 
$(T/\Lambda, \mu/\Lambda)=(0,1/e)$ at $g=g_{TCP}$
and that it approaches the $T$-axis asymptotically as $g$ getting larger; 
this fact proves that 
the TCP is only possible at finite number density in our model.

In this paper, for demonstration,
we set $g \Lambda^2=2.5$, and all other dimensionful quantities
will be measured in the unit of $\Lambda$. 
The phase diagram of the NJL model in the mean field approximation is
shown in Fig.~1. This shares the well-known structure of the
phase diagram possessing a tricritical point\cite{LS84}. The
character of this 
diagram around the TCP is studied using the Landau free energy and 
the Schwinger-Dyson equation
of the ladder QCD in Ref.\cite{HI03}. Their results on 
the quark mass dependence of the boundary curves
 and the behavior of the susceptibilities are 
consistent with those in NJL calculations\cite{FOO03}, because the
symmetry and order parameters of the system are the same.

\section{Response functions and meson spectrum}
\label{sec:3}
In order to clarify the structure of the relevant modes for
the phase transition, we should investigate the spectral 
functions\cite{HK85,KKKN01,PSS02,HMN02}.
Using the linear response theory, we first calculate the
response functions in the relevant channels, whose imaginary
parts give rise to the mode spectra.

Let us assume artificially
small deviations of the chemical potential
and the current quark mass, $\tilde \mu, \tilde m$,
from their equilibrium and physical values, respectively.
The response functions for these disturbances are calculated as
\begin{eqnarray}
\chi_{ab}(iq_4,{\bf q})
&=&
\Pi_{ab}(iq_4,{\bf q})+\Pi_{a m}(iq_4,{\bf q})
{1 \over 1-2g \Pi_{m m}(iq_4,{\bf q})} 2g \Pi_{m b}(iq_4,{\bf q}),
\qquad ( a,b = \mu, m).
\label{eq:response}
\end{eqnarray}
Here the polarizations  are defined with the 
imaginary--time quark
propagator ${\cal S}$ as
\begin{eqnarray}
\Pi_{mm}(iq_4,{\bf q})&=&
-\int {d^3 k \over (2 \pi)^3} T \sum_n 
tr_{\rm fcD} {\cal S}(\tilde k){\cal S}( \tilde k -q),
\\
\Pi_{\mu\mu}(iq_4,{\bf q})&=&
-\int {d^3 k \over (2 \pi)^3} T \sum_n 
tr_{\rm fcD} {\cal S}(\tilde k )i\gamma_4 {\cal S}(\tilde k -q)i\gamma_4 ,
\\
\Pi_{m\mu}(iq_4, {\bf q}) &=&
-\int {d^3 k  \over (2\pi)^3}T \sum_n  tr_{\rm fcD}
{\cal S}(\tilde k ){\cal S}(\tilde k -q)i \gamma_4,
\end{eqnarray}
where $q_4=2l\pi T$ $ (l\in \Z)$ is the Matsubara frequancy
of the bosonic mode, 
$\tilde k=({\bf k}, k_4 + i \mu)=
({\bf k}, -\omega_n + i \mu)$ is the quark loop momentum
with $\omega_n=(2n+1)\pi T$ $(n \in \Z)$,
and the traces are taken
over the flavor, color and Dirac indices.
The real--time response function is obtained by an analytic
continuation, which is simply achieved by a replacement
$iq_4 \to q_0 +i \epsilon$ after the Matsubara sum.
These response functions in the static case
should reduce to the
susceptibilities in Eq.~(\ref{eq:susceptibility}).
From this fact we find that the static 
polarization  $\Pi_{ ab}(0,\bq \to \bzero)$
are the free susceptibilities $\chi_{ab}^{(0)}(\bq \to \bzero)$,
which can be also  confirmed by explicit calculations.

The common denominator, ${1 \over 2g}- \Pi_{mm}(q_0,\bq)$, is proportional
to the inverse of the scalar response function, and its static
value coincides with the potential curvature $K$
in the long wave length limit.
Diagrammatically, this denominator corresponds to the sum of the bubble
diagrams in the scalar channel.
The gapless excitation mode which is responsible for the
appearance of the CEP and for all divergences of the scalar, vector and
entropy fluctuations through the coupling, must be contained
in the spectrum of this bubble diagram.

The only task we have to do, therefore, 
is to investigate the spectrum of the scalar
excitation determined by the condition 
\begin{equation}
\chi_{mm}^{-1}\propto
{1 \over 2g}- \Pi_{mm}(q_0,{\bf q}) =
{1 \over 2g}- J({\mathbf q})+(4M^2-q^2)I(q_0,{\bf q}) =0,
\label{eq:polecondition}
\end{equation}
where
\begin{eqnarray}
I(iq_4,{\bf q})&=&
\nu \int {d^3 k \over (2 \pi)^3}
T\sum_n {1 \over
(\tilde k^2+M^2)((\tilde k-q)^2+M^2)}
\nonumber \\
&=&
\nu \int^{k_{\rm max}} {d^3 k \over (2 \pi)^3}
{-1 \over 4 E_1 E_2}\left[
{1-n_{+1}-n_{-2} \over i q_4 -E_1-E_2}
-{n_{-1}-n_{-2} \over i q_4 +E_1-E_2}
+{n_{+1}-n_{+2} \over i q_4 -E_1+E_2}
-{1-n_{-1}-n_{+2} \over i q_4 +E_1+E_2}
\right ],
\label{eq:I}
\end{eqnarray}
and 
\begin{eqnarray}
J({\bf q})&=&
\nu \int {d^3 k \over (2 \pi)^3}
\left ( {1-n_{-1} - n_{+1} \over 2E_1}
+       {1-n_{-2} - n_{+2} \over 2E_2}\right  )
\nonumber \\
&=&
 {\nu \over 2 \pi^2 |{\bf q}|} \int_0^{k_{\rm max}} d k k
\left [ \omega_{\rm max}-\omega_{\rm min}
+ T \ln {(1+e^{-(\omega_{\rm max}-\mu)/T})
                (1+e^{-(\omega_{\rm max}+\mu)/T})
\over
                (1+e^{-(\omega_{\rm min}-\mu)/T})
                (1+e^{-(\omega_{\rm min}+\mu)/T})
}  \right ]
\end{eqnarray}
with $E_{1,2}=\sqrt{M^2+(\bk\pm \bq/2)^2}$,
$n_{\pm 1,2}=n_\pm(E_{1,2})$ and
$\omega_{\rm max,min}=\sqrt{M^2+\bk^2+\bq^2/4 \pm |\bq ||\bk |}$.
In these equations we shifted the three momentum ${\bf k}$ to 
make the integral more
symmetric and chose a simple convention of the cutoff for the 
mode with finite momentum $\bq$ 
as $k^2_{\rm max}=\Lambda^2-\bq^2/4$.
The stable mode should be obtained as 
a real solution of  Eq.~(\ref{eq:polecondition}).
The unstable, resonance mode will be found as a complex-valued
solution in the unphysical lower half plane of $q_0$.

The function $I(q_0,\bq)$ is elementary in the thermal field theory
\cite{LB96}.
By noting $1-n_+-n_- = (1-n_-)(1-n_+) - n_- n_+$ 
(pair creation $-$ annihilation) and
$n_{-1} - n_{-2} = (1-n_{-2})n_{-1}-(1-n_{-1})n_{-2}$
(emission $-$ absorption) and so on,
we find that the first and last terms in Eq.~(\ref{eq:I})
are contributions of the time--like spectrum
while
the second and third come from the space--like spectrum
in the spectral representation for $I(q_0,\bq)$.
It may be instructive to note here 
that two limits of the function $I(q_0,\bq)$ are
different: 
\begin{equation}
I(0,{\bf q} \to \bzero)=
\nu \int {d^3 k \over (2 \pi)^3}
{1 \over 4 E^2}\left(
{1-n_+-n_- \over E}
+n'_-+ n'_+
\right ),
\label{eq:space}
\end{equation}
\begin{equation}
I(q_0 \to 0,\bzero)=
\nu \int {d^3 k \over (2 \pi)^3}
{1 \over 4 E^2} {1-n_+  - n_-  \over E}.
\label{eq:time}
\end{equation}
The difference is due to the space-like mode contribution;
in the first limiting procedure 
$I$ has the space-like mode contribution
while in the second case we don't see it. 
This is the origin of the fact that 
the massive sigma meson and the zero curvature of the free
energy can cope with each other
[See also Eq.~(\ref{eq:curvature})].

\begin{figure}[tb]
\centerline{\epsfxsize=0.35\textwidth
\epsffile{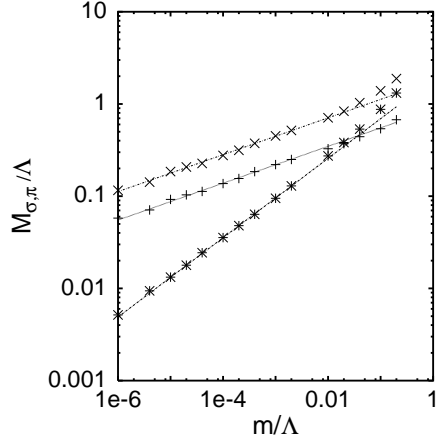}}
\caption{Sigma ($\times$) and pion ({\protect \STAR}) masses $M_{\sigma,\pi}$
at the CEP as
functions of the current quark mass $m$.
The constituent quark mass $M$ (+) is displayed as well.
\label{fig:sigmapi}}
\end{figure}

\subsection{Sigma and pion masses}
First we study the excitation energy of the massive mode,
which is continuously
connected to the sigma meson in the vacuum as the parameters $(T,\mu)$
are varied.
To this end we set $q=(q_0,{\bf q})=(\omega,\bzero )$,
following the usual calculation of the sigma meson mass in the
vacuum.  Then  
$J=\nu \int {d^3 k\over (2\pi)^3}(1-n_- -n_+)/E$,
and together with the gap equation
we recover
the well-known condition to determine the sigma energy, 
\begin{eqnarray}
&&{1\over 2g}{m \over M}+(4 M^2 - \omega ^2) I(\omega,\bzero )=0  .
\label{eq:sigma}
\end{eqnarray}
Similarly for the pions,
\begin{eqnarray}
&&{1\over 2g}{m \over M}- \omega^2 I(\omega,\bzero)=0 .
\label{eq:pi}
\end{eqnarray}
In the chiral limit
the energies of the stopped 
sigma and pion are $\omega=2M$ and 0, respectively,
which fact is known as Nambu relation\cite{NJL}
in the symmetry-broken phase.

At the CEP, on the other hand, the constituent quark mass
$M$ remains massive due to the non-zero current quark mass,
and therefore we expect the sigma meson
has a finite energy gap as a remnant of the Nambu relation.
From the argument using the Landau free energy, 
$M$ is expected to scale with $m^{1/5}$ at the CEP
\cite{HI03,LS84}. 
With neglecting the imaginary part of $I$,
we numerically confirm
this behavior,
in Fig.~\ref{fig:sigmapi}, 
 for the sigma mass $M_\sigma$ determined by Eq.~(\ref{eq:sigma})
as well as the constituent quark mass $M$.
The pion mass will be proportional to $(m/M)^{1/2} \sim m^{2/5}$
from Eq.~(\ref{eq:pi}), and
it is numerically found as $\sim m^{0.43}$.
It should be stressed here that the massive behavior of the sigma is
quite normal from the viewpoint of the Nambu relation,
and  does not introduce any problem in the description of 
 the phase transition.
Actually it is found that 
the massive sigma and the vanishing curvature take place at the same time
at the CEP in the NJL
 model \cite{OO02}, as well as in the linear sigma model\cite{F02},
in the mean--field approximation.

\begin{figure}
\centerline{\epsfxsize=0.48\textwidth \epsffile{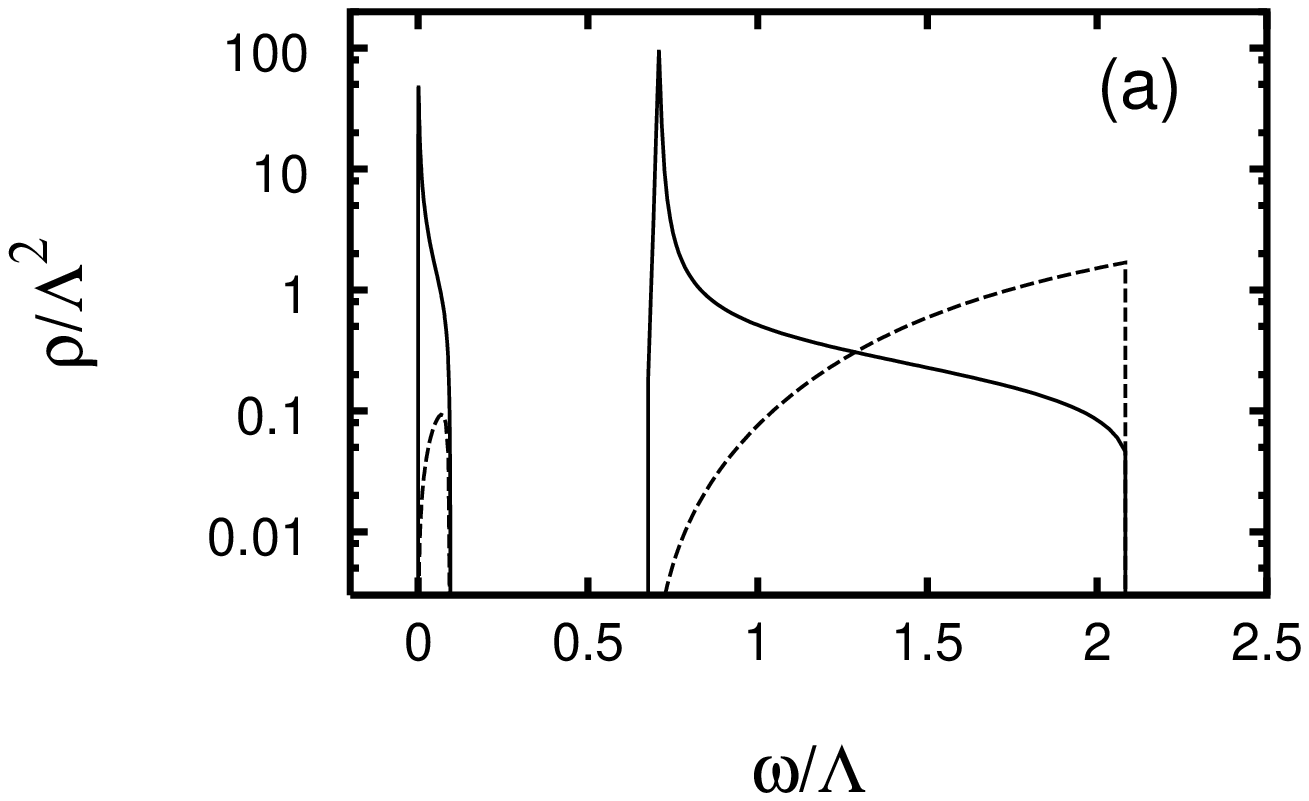}
\epsfxsize=0.48\textwidth \epsffile{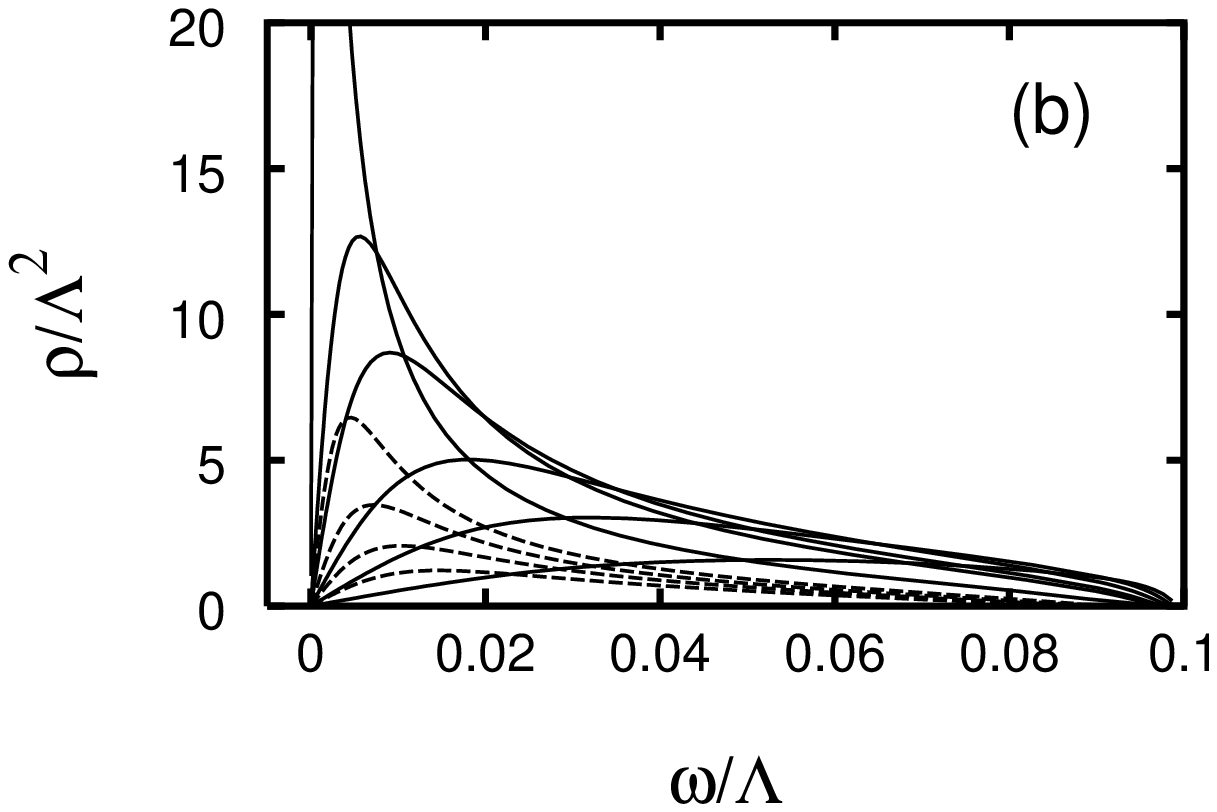}}
\caption{Calculated spectral function in the scalar mode with
$(T/\Lambda,m/\Lambda, |\bq|/\Lambda) =(0.14985,0.01,0.1)$.
(a) Profile at the CEP (solid line) together
with the free spectrum (dashed line).
(b) Profile in the space--line region
with varying the chemical potential as
$\mu/\Lambda$
=0.59, 0.58, 0.575, 0.572, 0.571, 
 0.57006 ($=\mu_c$) (solid lines),
0.5682, 0.5652, 0.56, 0.55 (dashed lines).
\label{fig:sp}}
\end{figure}

\subsection{Spectrum of the scalar mode and the phase transition}

As we explained in Introduction, we expect that the instability
occurs in the almost static and long wave length region in the scalar
density fluctuation
(as well as in the vector and entropy fluctuations through the coupling).
In order to investigate the spectrum of the relevant modes
in the transition at the CEP, we must keep the momentum ${\bf q}$
finite.

The spectral function of the scalar mode, 
$\rho(\omega,\bq)= 2 {\rm Im} \chi_{mm}(\omega ,\bq)$, 
is explicitly written as
\begin{equation}
{1\over 2}\rho(\omega,\bq)
={\rm Im} \chi_{mm}(\omega, {\bf q})=
{-({1 \over 2g })^2 (4M^2-q^2) {\rm Im} I(\omega, {\bf q}) \over 
[{1 \over 2g}-J(\bq)+(4 M^2-q^2){\rm Re}I(\omega,{\bf q})]^2+
[(4M^2-q^2){\rm Im} I(\omega, {\bf q})]^2 }
\label{eq:spectrum}
\end{equation}
up to delta functions coming from possible non-decaying stable modes. 
We noticed previously 
that $\Pi_{mm}(\omega, \bq)$
is the scalar response function of  the
free quark gas of mass $M$, 
whose imaginary part
\begin{equation}
{1\over 2}\rho^{(0)}(\omega,\bq)=
{\rm Im} \Pi_{mm}(\omega, {\bf q})=
-(4M^2-q^2) {\rm Im} I(\omega, {\bf q})
\end{equation}
is the scalar channel spectrum of the uncorrelated quarks.
These spectral functions have
the support in the time-like region
with $q^2=\omega ^2-{\bf q}^2 > 4M^2$ and in the space-like region 
with $q^2=\omega ^2-{\bf q}^2 <0$.

In the numerical calculation presented here,
we fix the current quark mass to $m/\Lambda =0.01$
for definiteness.
The corresponding CEP locates at $
(T/\Lambda, \mu/\Lambda)= (0.14985,0.57006)$.
The spectrum of $\chi_{mm}(\omega, \bq)$ with
$|\bq|/\Lambda = 0.1$ is shown in Fig.~\ref{fig:sp} (a),
in which we see two sharp peaks at very low frequency
and around $2 M$ ($M/\Lambda=0.331$).
The free spectrum is also displayed for comparison.
It is understood that the
scalar channel attraction, which is encoded as a bubble sum
in the denominator in Eq.~(\ref{eq:response}) or 
in Eq.~(\ref{eq:spectrum}),
changes the structure
of the spectrum dramatically.
The resulting scalar mode develops
two distinct branches.
The higher frequency peak corresponds to the sigma meson while
the lower one appearing in the space--like region 
($\omega < |\bq|$) is the scalar
density fluctuating mode which we expect is responsible for the second
order phase transition.
Actually the latter peak diverges as $\bq \to \bzero$. 
We study  the structure of the peak in the space--like 
region near the CEP in more detail in Fig.~\ref{fig:sp} (b).
The enhancement of the spectrum is clearly
demonstrated as the CEP is approached from the higher densities,
$\mu/\Lambda =0.59, 0.58, 0.575, 0.572, 0.571$ (solid lines), and
also from the lower $\mu/\Lambda =0.55, 0.56, 0.5652, 0.5682$
(dashed lines) with $\mu_c/\Lambda=0.57006$.

\begin{figure}[tb]
\centerline{\epsfxsize=0.48\textwidth \epsffile{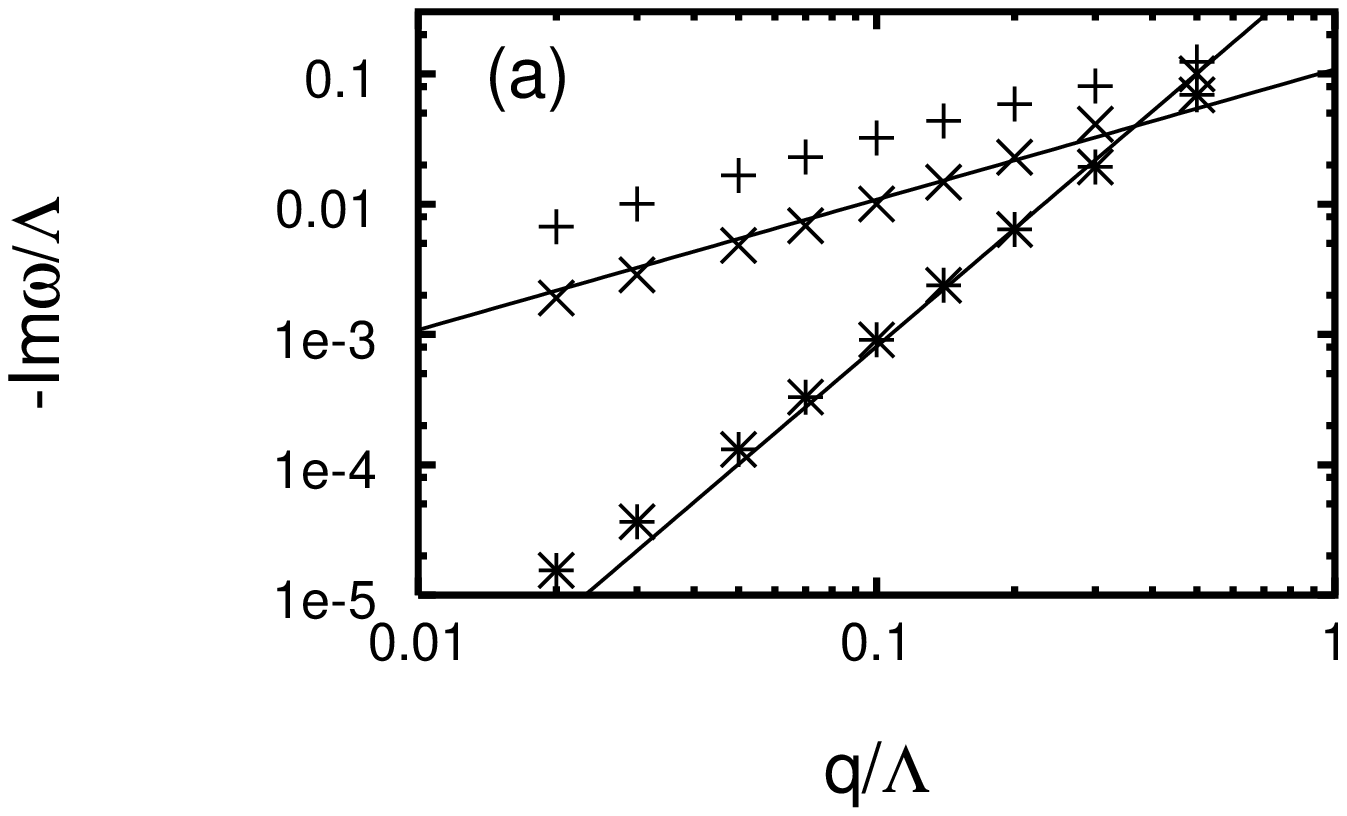}
\epsfxsize=0.48\textwidth \epsffile{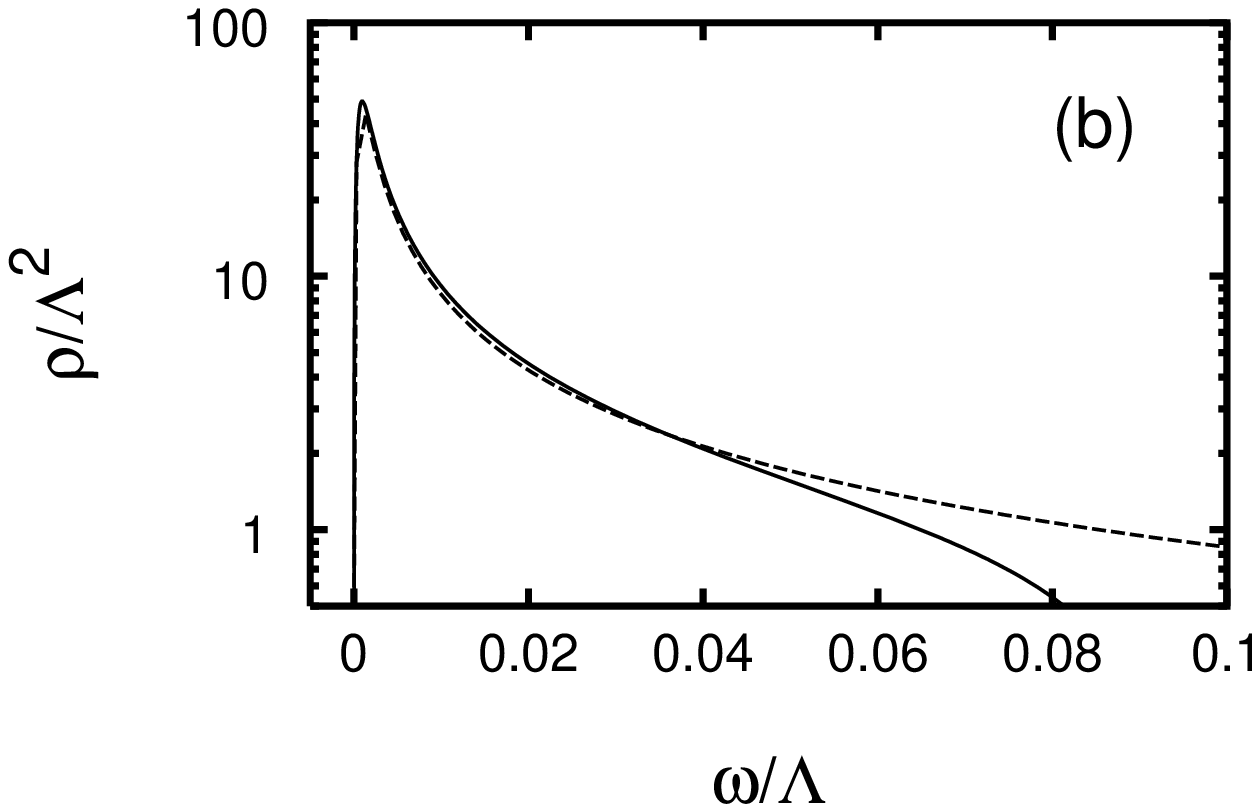}}
\caption{(a) The pole position of $\chi_{mm}(\omega,\bq)$
 as a function of $|\bq|/\Lambda$
for the CEP ({\protect \STAR}), a lower density ($\mu/\Lambda=0.56$, $\times$),
and a higher density ($\mu/\Lambda=0.58$, +), together
with the 
straight lines ($\propto |\bq|^3$ and $|\bq|^1$).
(b) Spectral function of the single pole form Eq.~(\ref{eq:param})
(dashed line) in the 
space-like region at the CEP is compared with the numerical result
(\ref{eq:spectrum}) (solid line)
with $|\bq|/\Lambda =0.1$.
\label{fig:singlepole}}
\end{figure}

We searched the pole in the unphysical lower half plane of
$\omega$\cite{KKKN01,PSS02,HMN02},
responsible for this peak.
Indeed the pole is found on the
negative imaginary axis as shown in Fig.~\ref{fig:singlepole} (a).
We observed that the pole is moving toward
the origin as decreasing the momentum $\bq$.  
Even in the non-critical case there is a pole on the negative imaginary
axis and it moves toward the origin as $\bq \to \bzero$, too.
The difference between the critical and non-critical cases exists in
the exponent of the $\bq$-dependence;
the pole position goes to zero as $|\bq|^3$ in the critical case
while  on the other hand it behaves like $|\bq|^1$ for non-critical case.

This result can be understood if one notes that in the static limit
 of the ``sound'' mode,
$\omega \to 0$ with $u\equiv \omega /|\bq|<1$ fixed,
the function $I$ has the form ($v=k/E$) \cite{FW71}
\begin{equation}
I(u)=I(0,|\bq|\to 0) - {\nu \over 2\pi^2} \int_0^\Lambda dk 
(n'_- + n'_+) {uv \over 8}\left (\ln \left |{v+u \over v-u} \right | 
-{\rm i}\pi \theta (1-{u \over v})\right ),
\end{equation}
whose imaginary part is proportional to $u=\omega/|\bq|$ for small $u$.
This imaginary part physically expresses the damping of 
the scalar ``sound'' mode in the quark gas.
Then the response function with the space-like kinematics
near the CEP 
may be approximated with
\begin{equation}
\chi_{mm}(\omega, {\bf q}) 
\sim 
{{\rm Re}\Pi_{mm}(0,\bq) \over - {\rm i}2g {\rm Im}\Pi_{mm}(\omega,\bq) 
+(1-2g {\rm Re}\Pi_{mm}(0,\bq))}
={1 \over -{\rm i} {\omega \over \lambda({\bf q})}+\chi^{-1}_{mm}({\bf q})}
= {\lambda({\bf q}) \over -{ \rm i} \omega+\omega_c(\bq)} ,
\label{eq:param}
\end{equation}
where $\chi_{mm}(\bq)$
is the scalar susceptibility,
and $\omega_c (\bq) \equiv \chi_{mm}^{-1}({\bf q})\lambda({\bf q})$.
Here  we  expect $\lambda(\bq) \propto |\bq|$, and numerically confirmed it.
The susceptibility behaves as 
$\chi^{-1}_{mm}(\bq\to \bzero) \propto |\bq|^2$ 
at the CEP, but stays constant in non-critical region.
This difference of the $\bq$-dependence of $\chi_{mm}(\bq)$
alters the exponent of the $\bq$-dependence of the pole position,
as described above and shown in Fig.~\ref{fig:singlepole} (a).

We confirmed in Fig.~\ref{fig:singlepole} (b) that
the $\omega$-dependence of the
response function at the CEP is well reproduced in the single pole form
(\ref{eq:param}) with the values 
$(\omega_c(\bq),\chi^{-1}_{mm}(\bq))
=(0.90925\cdot 10^{-3},0.21300\cdot 10^{-1})$
obtained in our numerical pole search.
The deviation seen around $\omega/\Lambda \sim 0.1$ is due to the
lack of the kinematical cutoff to confine the spectrum inside the
space-like region when we use the single pole form (\ref{eq:param}).
We also checked that the response functions
for the non-critical case near the CEP
can be also described with the single pole form fairly well.
Incidentally it is immediately seen that the form of (\ref{eq:param})
has the different limits around the origin:
if one take the $\bq \to \bzero$ limit first, 
$\chi_{mm}(\omega, 0) \to 0$.
On the other hand, 
$\chi_{mm}(0, \bq) = \chi_{mm}(\bq) \to \infty$
as $\bq \to \bzero$.

The typical slow frequency of the system near the CEP
is $\omega_c(\bq) =
 \chi_{mm}^{-1}(\bq)\lambda(\bq)$.
This means that we reconfirmed
 the conventional theory of dynamical critical phenomena
\cite{LVH54} in the form of Eq.~(\ref{eq:param}):
the divergence of the susceptibility is directly related to the
critical slowing down of the response of the system 
to the external disturbance.
The exponent of the dynamical scaling in our model will be equal to 3.

\section{Summary}
\label{sec:4}
In this paper we have explicitly shown that there are
two branches in the scalar mode in the NJL model at
finite ($T, \mu$).
The sigma meson mode stays massive at the CEP, following
the approximate relation, $M_\sigma \sim 2M$.
The other mode is the scalar density fluctuation
which becomes soft and relevant
near the CEP.
This result is perfectly consistent 
with the description of the second order phase transition
using the Landau free energy.
The first observation of the massive sigma at the CEP was done
in Ref.~\cite{SMMR01}
although it led the authors to a somewhat confused discussion.

All the susceptibilities $\chi_{ab} (a,b=T,\mu,m)$ 
are found divergent at the CEP due to the flatness of the potential
curvature, $K\to 0$. The importance of the
density fluctuation at the CEP is shown by considering the number-density-fixed
susceptibility, $\chi_{mm}(T,\rho_q)$. It no longer diverges
at the CEP, but still does at the chiral phase transition.
This situation at the CEP is quite similar to the
liquid--gas phase transition, where the density difference 
between two phases is identified as the relevant
order parameter.

From our analysis we conclude that 
the CEP line in Fig.~\ref{fig:phase} denotes the
second order phase transition characterized by the instability
of the density fluctuations. 
Two lines of the
CEP and the line of the second order chiral transition meet
at the TCP (therefore tricritical),
where we expect that the sigma meson and the 
density fluctuation mode become unstable at
the same time. Although 
some readers could expect our conclusion from the
beginning,
there are confusions on this point in the 
literature \cite{RS98,SMMR01,KF03}
and one of our contributions
is to clarify the importance of the density fluctuation
explicitly using
a concrete model for the CEP of a quark system.

As for the experimental signatures of the CEP one should investigate
first the implications of the anomalous fluctuations
 \cite{RS99,SG99,HI03,HS03,AHM00,JK00,BG01} of
the baryon number, entropy, and scalar densities.
One should be careful that the modification of the density fluctuations
with the space--like momentum
does not affect directly the spectrum of the
particle production modes, 
like $\sigma \to \pi \pi, \bar ll, \gamma\gamma$, 
because of the kinematical mismatching.

\acknowledgments
The author expresses his sincere thanks to K.~Fukushima,
T.~Matsui, O.~Morimatsu, K.~Ohnishi, M.~Ohtani for fruitful
discussions and to the members of Komaba nuclear theory group
for their interests on this subject.
This work is supported in part by the Grants-in-Aid for Scientific
Research of Monka-sho (Grant No.~13440067).


\begin{thebibliography}{99}
\bibitem{AY89} M.~Asakawa and K.~Yazaki, Nucl.~Phys.~{\bf A504} 668,1989.
\bibitem{GGP94}
S.~Gavin, A.~Gocksch, R.D.~Pisarski, Phys.~Rev.~D {\bf 49}
3079, 1994 (hep-ph/9311350). 
\bibitem{BR99} J.~Berges and K.~Rajagopal, Nucl.~Phys.~{\bf B538} 215,
  1999 (hep-ph/9804233).
\bibitem{HJSSV98} M.A.~Halasz, A.D.~Jackson, R.E.~Shrock,
  M.A.~Stephanov, and J.J.M.~Verbaarschot, Phys.~Rev.~D {\bf 58} 
096007, 1998 (hep-ph/9804290).
\bibitem{FK02} Z.~Fodor and S.D.~Katz, JHEP 0203, 014,2002 (hep-lat/0106002).
\bibitem{RS98}
M.A.~Stephanov, K.~Rajagopal and  E.V.~Shuryak, 
Phys.~Rev.~Lett. {\bf 81} 4816,1998 (hep-ph/9806219).
\bibitem{RS99}
M.A.~Stephanov, K.~Rajagopal and E.V.~Shuryak, 
Phys.~Rev.~D {\bf 60} 114028,1999 (hep-ph/9903292). 
\bibitem{SG99} S.~Gavin, nucl-th/9908070.
\bibitem{BR00} B.~Berdnikov and K.~Rajagopal, Phys.~Rev.~D {\bf 61}
105017, 2000.
\bibitem{SMMR01} O.~Scavenius, A.~Mocsy, I.N.~Mishustin and D.H.~Rischke,
Phys.~Rev.~C {\bf 64} 045202, 2001 (nucl-th/0007030). 
\bibitem{BPSS01}S.~Borsanyi, A.~Patkos, D.~Sexty, Z.~Szep,
Phys.~Rev.~D {\bf 64} 125011, 2001 (hep-ph/0105332).
\bibitem{HI03} Y.~Hatta and T.~Ikeda, Phys.~Rev.~D {\bf 67} 014028,
  2003 (hep-ph/0210284).
\bibitem{KF03} K.~Fukushima, Phys.~Rev.~C {\bf 67} 025203 (hep-ph/0209270).
\bibitem{HS03} Y.~Hatta and M.A.~Stephanov, hep-ph/0302002. 
\bibitem{HK94} T.~Hatsuda and T.~Kunihiro, Phys.~Rept. {\bf 247} 221, 1994.
\bibitem{SPK92} S.P.~Klevansky Rev.~Mod.~Phys. {\bf 64} 649, 1992.
\bibitem{HH77} P.C.~Hohenberg and B.I.~Halperin,
  Rev.~Mod.~Phys. {\bf 49} 435, 1977.
\bibitem{FOO03} H.~Fujii, K.~Ohnishi, and M.~Ohtani,
 in preparation.
\bibitem{NJL} Y.~Nambu and G.~Jona-Lasinio, Phys.~Rev. {\bf 122} 345, 1961;
{\bf 124} 246, 1961.
\bibitem{TK91} T.~Kunihiro, Phys.~Lett.\ {\bf B271} 395, 1991.
\bibitem{AG91} A.~Gocksch,  Phys.~Rev.~Lett.\ {\bf 67} 1701, 1991.
\bibitem{HES77} H.E.~Stanley, {\em Introduction to phase transitions
and critical phenomena}, Oxford (1977).
\bibitem{LS84} I.~Lawrie and S.~Sarbach, in {\em Phase Transitions and
    Critical Phenomena}, ed.~by C.~Domb and J.~Lebowitz (Academic
  Press, NY, 1984), Vol.~9, pp.~1.
\bibitem{HK85} T.~Hatsuda and T.~Kunihiro, Phys.~Rev.~Lett.{\bf 55}
  158, 1985.
\bibitem{KKKN01} M.~Kitazawa, T.~Koide, T.~Kunihiro and Y.~Nemoto
Phys.~Rev.~D {\bf 65} 091504(R), 2002 (nucl-th/0111022).
\bibitem{PSS02} A.~Patkos, Z.~Szep and P.~Szepfalusy, Phys.~Rev.~D {\bf 66} 
116004, 2002 (hep-ph/0206040).
\bibitem{HMN02} Y.~Hidaka, O.~Morimatsu and T.~Nishikawa, hep-ph/0211015.
\bibitem{LB96} M.~Le Bellac, {\em Thermal Field Theory}, Cambridge
University Press (1996).
\bibitem{OO02} K.~Ohnishi and M.~Ohtani, private communications.
\bibitem{F02} K.~Fukushima, private communications.
\bibitem{FW71} A.L.~Fetter and J.D.~Walecka, 
{\em Quantum Theory of Many--Particle Systems}, 
McGraw Hill (1971).
\bibitem{LVH54} L.~van Hove, Phys.~Rev.~{\bf 95} 249;1374, 1954.
\bibitem{AHM00} M.~Asakawa, U.~Heinz and B.~Muller,
Phys.~Rev.~Lett.\ {\bf  85} 2072, 2000 (hep-ph/0003169).
\bibitem{JK00} S.~Jeon and V.~Koch, Phys.~Rev.~Lett. {\bf 85} 2076, 
2000 (hep-ph/0003168).
\bibitem{BG01} D.~Bower and S.~Gavin, Phys.~Rev.~C {\bf 64}
051902, 2001 (nucl-th/0106010). 
\end{thebibliography}
\end{document}